\documentclass{llncs}

\usepackage{graphicx}

\usepackage{authblk}

\usepackage{algorithm}

\usepackage{algorithmic}

\usepackage[figuresright]{rotating} 

\usepackage{epsfig,times,latexsym,enumerate,lscape,flafter,amsmath,amssymb,graphicx}

\begin{document}

\title{An investigation of the classifiers to detect android malicious apps}

\titlerunning{An investigation of the classifiers to detect android malicious apps}

\author{Ashu Sharma \and Sanjay Kumar Sahay\inst{*}}

\authorrunning{Ashu Sharma et al.}

\institute{Department of Computer Science and Information System, Birla Institute
of Technology and Science, K. K. Birla Goa Campus, NH-17B, By Pass Road, Zuarinagar-403726, Goa, India
\email{\{p2012011, ssahay\}@goa.bits-pilani.ac.in}}

\maketitle

\begin{abstract}

Android devices are growing exponentially and are connected through the internet accessing billion of online websites. The popularity of these devices encourages malware developer to penetrate the market with malicious apps to annoy and disrupt the victim. Although, for the detection of malicious apps different approaches are discussed. However, proposed approaches are not suffice to detect the advanced malware to limit/prevent the damages. In this, very few approaches are based on opcode occurrence to classify the malicious apps. Therefore, this paper investigates the five classifiers using opcodes occurrence as the prominent features for the detection of malicious apps. For the analysis, we use WEKA tool and found that FT detection accuracy ($\sim$79.27\%) is best among the investigated classifiers. However, true positives rate i.e. malware detection rate is highest ($\sim$99.91\%) by RF and fluctuate least with the different number of prominent features compared to other studied classifiers. The analysis shows that overall accuracy is majorly affected by the false positives of the classifier.
\end{abstract}
\section{Introduction}\label{sec:Introduction}

Android is one of the most popular operating system in for smart devices and are connected through the internet accessing billions of online websites. The exponential increase in android apps is basically due to the open source, third party distribution, free rich SDK and the very much suited java language. 
In this growing android apps market, it is very hard to know which apps are  spam or malware content. As per statista \cite{statista} $\sim2 \times 10^6$ android apps are available at google play store. Also, there are many third party android apps available for the users \cite{9apps}, which may be malicious. Hence potential of the malicious apps or malware entering these systems is now at never seen before levels. 
\par Due to ease of use, these devices  hold sensitive information such as personal data, browsing history, shopping history, financial details, etc. \cite{qr2015} i.e. users are ever more frequent to use the internet consequently these devices are vulnerable to cyber threats/attacks. In this, Quick Heal Threat Research Labs in the 3rd quarter of 2015 reported that they have received samples of files at the rate of $\sim4.2 \times 10^5$
samples per day for the Android and Windows platforms and the G Data security experts expect a rapid increase in numbers of new malware samples in 2016 compare to previous years \cite{gdata2015}.

The traditional approach i.e. signature based techniques, to detect the advanced malicious android apps are no longer effective, as it uses code obfuscation techniques. However, a number of methods have been proposed on static and dynamic analysis for analyzing and detecting Android malware prior to their installation \cite{enck2014taintdroid} \cite{felt2011android} \cite{grace2012riskranker}  \cite{reina2013system}  \cite{yan2012droidscope}. 
It appears that so far proposed approaches are not suffice to detect the advanced malware to limit/prevent the damages \cite{sharma2014evolution}. 
Therefore, we investigated the five classifiers ( FT, Random forest, J48, LMT and NBT ) and present a novel approach to combat malware threat/attack by analysing the opcode occurrence in the apps. 
The remaining paper is organised as follows. In next section, we discuss the related work. Section 3 describe our approach to detect the malicious apps based on static analysis. The results of our approach are discussed in section 4. Finally, section 5 contains the conclusion and direction for the future work.

\section{Related work}\label{sec:Related work}
 
Static and dynamic analysis are the two main approaches applied for detection of android malware \cite{sharma2014evolution}. In static analysis, without executing the apps, the code are analysed to find a malicious pattern by extracting the features such as permissions, APIs used, control flow, data flow, broadcast receivers, intents, hardware components etc. Whereas, in the dynamic analysis the apps are examined in run time environment by monitoring the dynamic behaviour (network connections, system calls, resources usage, etc.) of the apps and the system response. However, in both the approaches selected classifiers are trained with a known dataset to differentiate the benign and malicious apps.
In this Seo, et. al. by analysing  the permissions, dangerous APIs and keywords associated with malicious behaviours detected potential malicious scripts in Android apps \cite{seo2014detecting}.
A lightweight framework was discussed by Arp, et. al., which uses AndroidManifest.xml file and disassembled code to generate a joint vector space \cite{arp2014drebin}.
Wu, et. al., approach detects the malware by analyzing AndroidManifest.xml and tracing the systems calls \cite{wu2012droidmat}.
Sanz, et. al., analysed five classifiers with machine learning (DT, KNN, BN, RF \& SVM) for automatic malware detection by analysing different sets of Android market permissions, ratings and a number of ratings. They found that among five classifiers BN performs the best while RF second and DT worst \cite{sanz2012automatic}.
Vidas, et. al., developed a tool which automatically analyzes the apps to find the least permissions/privileges that are required to run the apps \cite{vidas2011curbing}.
In this, Fuchs, et. al., method analyse the data flow across the android apps components \cite{fuchs2009scandroid}.
Daniel, et. al., did a broad static analysis by embedding the features in a joint vector space, such that the typical patterns of malware can be automatically identified \cite{arp2014drebin}. 
In the DREBIN project, a study has been done with 123,453 benign and 5,560 malware apps.
 Based on a set of characteristics derived from binary and metadata Gonzalez, et. al., proposed a method named as DroidKin, which can detect the similarity among the apps under various levels of obfuscation (code reordering, register reassignment, etc. \cite{sharma2014evolution} \cite{sharma2016improving}) \cite{gonzalez2014droidkin}.
SVM-based malware detection scheme given by Gugian, et. al., integrates both risky permission combinations and vulnerable API calls and used them as features for the classification \cite{scholkopf2001estimating}. 
Saracino, et. al., 2016 \cite{saracino2016madam} proposed a novel host-based malware detection system called MADAM which simultaneously analyzes and correlates the features at four levels (kernel, application, user and package) to detect and stop the malicious behaviours. Quentin et. al., uses op-code sequences to detect the malicious apps, however the approach will not detect completely different malware \cite{jerome2014using}.
Later on using N-opcode, BooJoong et. al., classified the malware and reported F-measure 98\% \cite{kang2016n}.

\section{Our approach}\label{sec:Our approach}

\begin{figure}[!htb]
    \centering
       \includegraphics[width=0.75\linewidth, height=0.20\textheight]{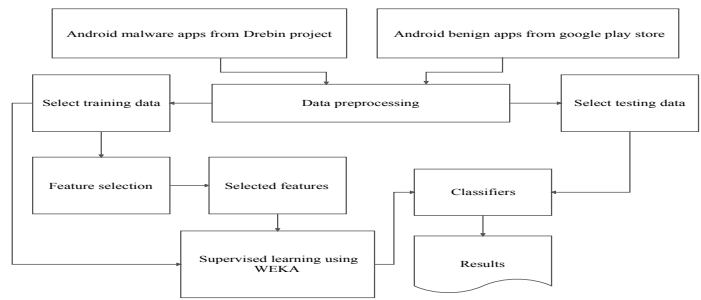} 
\caption{\small \sl Flow chart of the proposed approach for detection of android malicious apps.} 
\label{fig:fc}
  \end{figure}

  A novel approach to classify the unknown android malware is shown in figure~\ref{fig:fc}, which involves finding the promising features (algorithm.~\ref{algo:FS}), classifiers training and its detection.

  \vspace{-0.3in}
  
\subsection{Data Preprocessing and Feature Selection}
For the classification of unknown android malware apps, we downloaded 5531 android malware from DREBIN \cite{arp2014drebin} and 2691 benign apps from google play store. The benign apps are cross verified from virustotal.com \cite{wp1}.
\par
\begin{figure}[!htb]
    \centering
\includegraphics[width=0.95\linewidth, height=0.20\textheight]{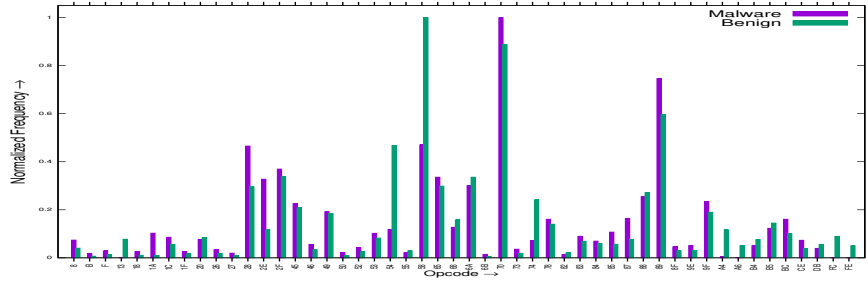} 
\caption{\small \sl Dominant opcodes of malicious and benign android apps .} 
\label{fig:f1}
  \end{figure}

 To understand the logic of android malware apps, we use freely available \textit{apktool} \cite{winsniewski2012android} to decompress the android $.apk$ files. After decompressing, we kept $.smali$ files and discarded other created files/folders. The $.smali$ files contains only one class information and is equivalent to $.class$ file.
To find the prominent features for classification of android malware and benign, we extracted the opcodes (list of the android opcodes is available at http://pallergabor.uw.hu/androidblog/dalvik\_opcodes.html) of the apps from the obtained $.smali$ files.
We analysed the opcode occurrence of all the android apps and found that the occurrence of many opcodes in malware and benign apps differ in large.
The normalized opcode occurrence of both the apps are shown in figure. \ref{fig:f1}. The mapping of the opcodes with hexadecimal representation has been kept same as given by the android developers \cite{opcodeList}. 
The prominent opcodes (features), which suppose to distinguish the malicious and benign android apps are obtained as described in the algorithm.~\ref{algo:FS}.
For the classification, we have used Waikato Environment for Knowledge Analysis (WEKA) tool, a collection of visualisation tools and algorithms for data analysis and predictive modeling, together with graphical user interfaces for easy access to this functionality \cite{holmes1994weka}, in which many inbuilt classifiers are available. 
On the  basis of studies done by Sharma and Sahay \cite{sharma2016effective} \cite{sahay2016grouping}, we selected the best classifier (Random forest \cite{rodriguez2006rotation}, LMT (Logistic model trees) \cite{landwehr2005logistic}, NBT (Naive-Bayes tree) \cite{kohavi1996scaling}, J48 \cite{bhargava2013decision} and FT (Functional Tree) \cite{gama2004functional}) for in-depth analysis by using K-fold cross-validation technique.

 \vspace{-0.2in}
 
 \vspace{-0.4in}

\begin{algorithm}[!htb]
\textbf{INPUT:} Pre-processed data\\
$\mathbf{N_B}$:  Number of benign android apps, $\mathbf{N_M}$: Number of malware android apps, 
\\$\mathbf{n}$: Total number of prominent features required.
\\ \textbf{OUTPUT:} List of prominent features
\begin{algorithmic}

\STATE \textbf{BEGIN}

\FORALL{benign apps }
\STATE Compute sum of the frequencies $\mathbf{f_i}$ of each opcode $\mathbf{Op}$  and normalize it. 
\STATE \begin{equation*} 
F_B ( Op_j ) = ( \sum f_i ( Op_j ) ) / N_B  
\end{equation*}
\ENDFOR

\FORALL{malware data }
\STATE Compute sum of the frequencies $\mathbf{f_i}$ of each opcode $\mathbf{Op}$  and normalize it.
\STATE \begin{equation*} 
F_M ( Op_j ) = ( \sum f_i ( Op_j ) ) / N_M 
\end{equation*}
\ENDFOR

\FORALL{opcode $\mathbf{Op_j}$}
\STATE Find the difference of the normalized frequencies for each opcode $\mathbf{D(Op_j)}$.
\STATE \begin{equation*} 
D(Op_j)= | F_B ( Op_j ) - F_M ( Op_j ) |
\end{equation*}
\ENDFOR

\RETURN $\mathbf{n}$ number of prominent opcodes as features with high $\mathbf{D(Op)}$.
\end{algorithmic}
\caption{\textbf{:} Feature Selection}
\label{algo:FS}
\end{algorithm}

\begin{figure}[!htb] 
\centering 
\begin{minipage}[t]{0.49\textwidth}
\includegraphics[width=1\linewidth, height=0.3\textheight]{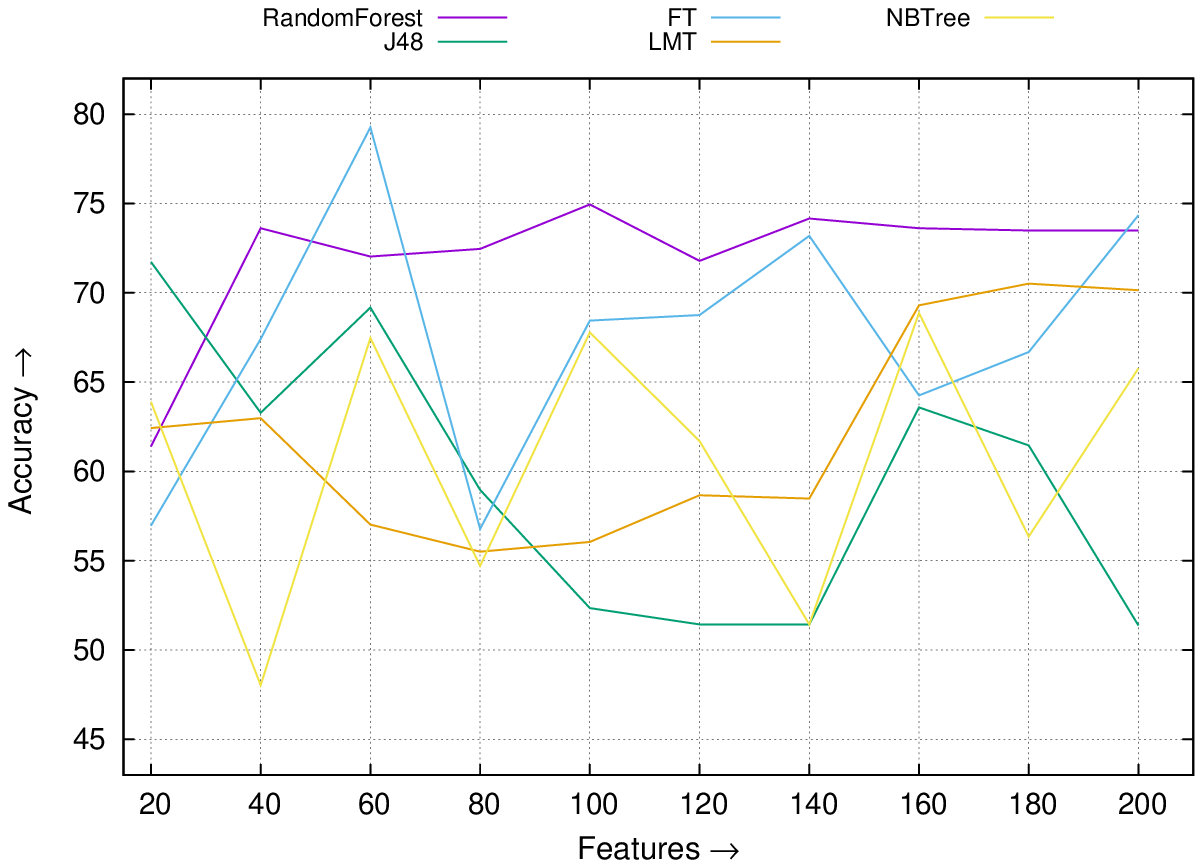} 
\caption{\small \sl Detection accuracy obtained by the selected five classifiers with different number of prominent features.} 
\label{fig:accuracy}
  \end{minipage}  
\begin{minipage}[t]{0.49\textwidth}
       \centering
     \includegraphics[width=1\linewidth, height=0.3\textheight]{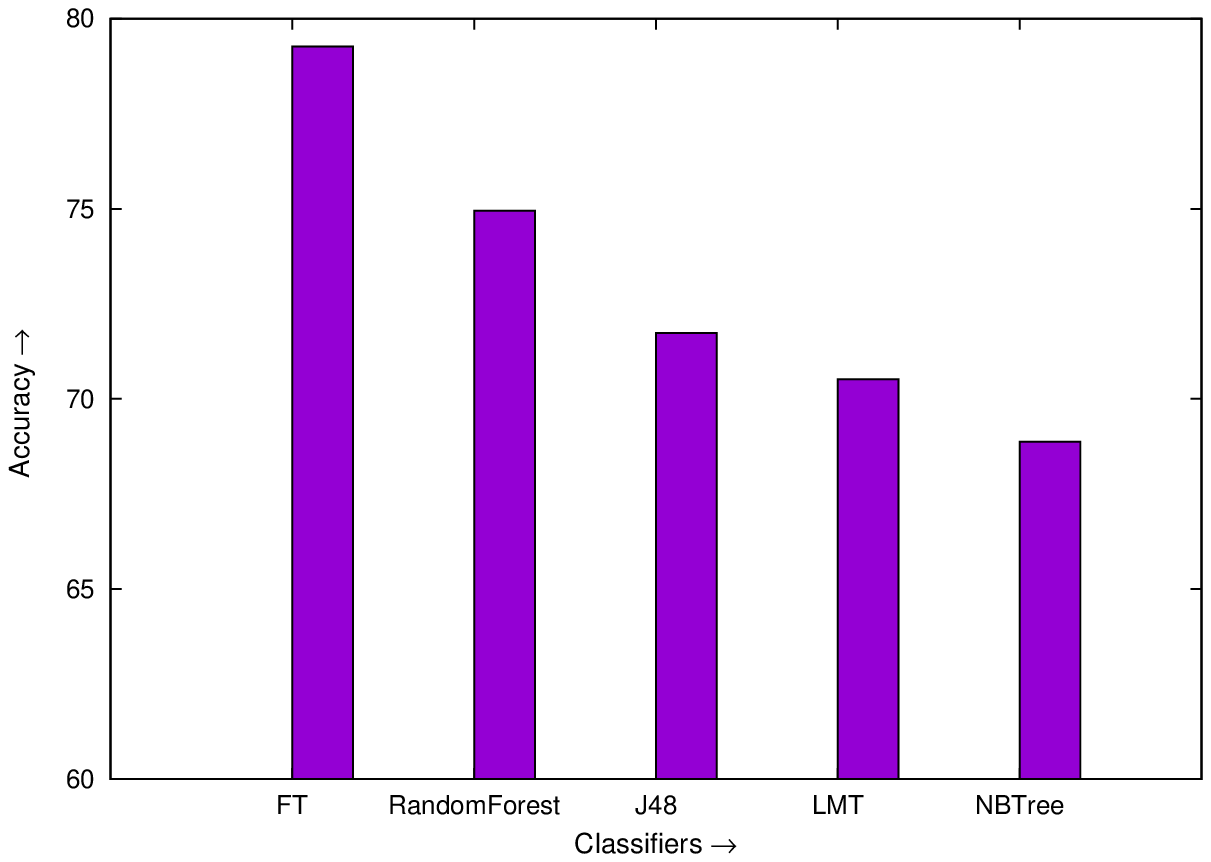} 
\caption{\small \sl Best accuracy obtained by the selected five classifiers.} 
\label{ba}
    \end{minipage}   
\end{figure}

\begin{figure}[!htb] 
\centering 
\begin{minipage}[t]{0.49\textwidth}
\centering
        \includegraphics[width=1\linewidth, height=0.3\textheight]{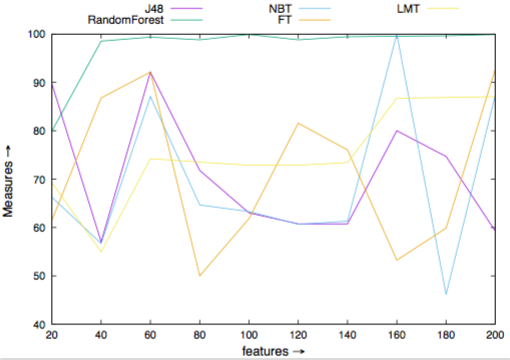} 
\caption{\small \sl True positives obtained by selected five classifiers with different number of prominent features.} 
\label{tp}
  \end{minipage}  
\begin{minipage}[t]{0.49\textwidth}
       \centering
      \includegraphics[width=1\linewidth, height=0.3\textheight]{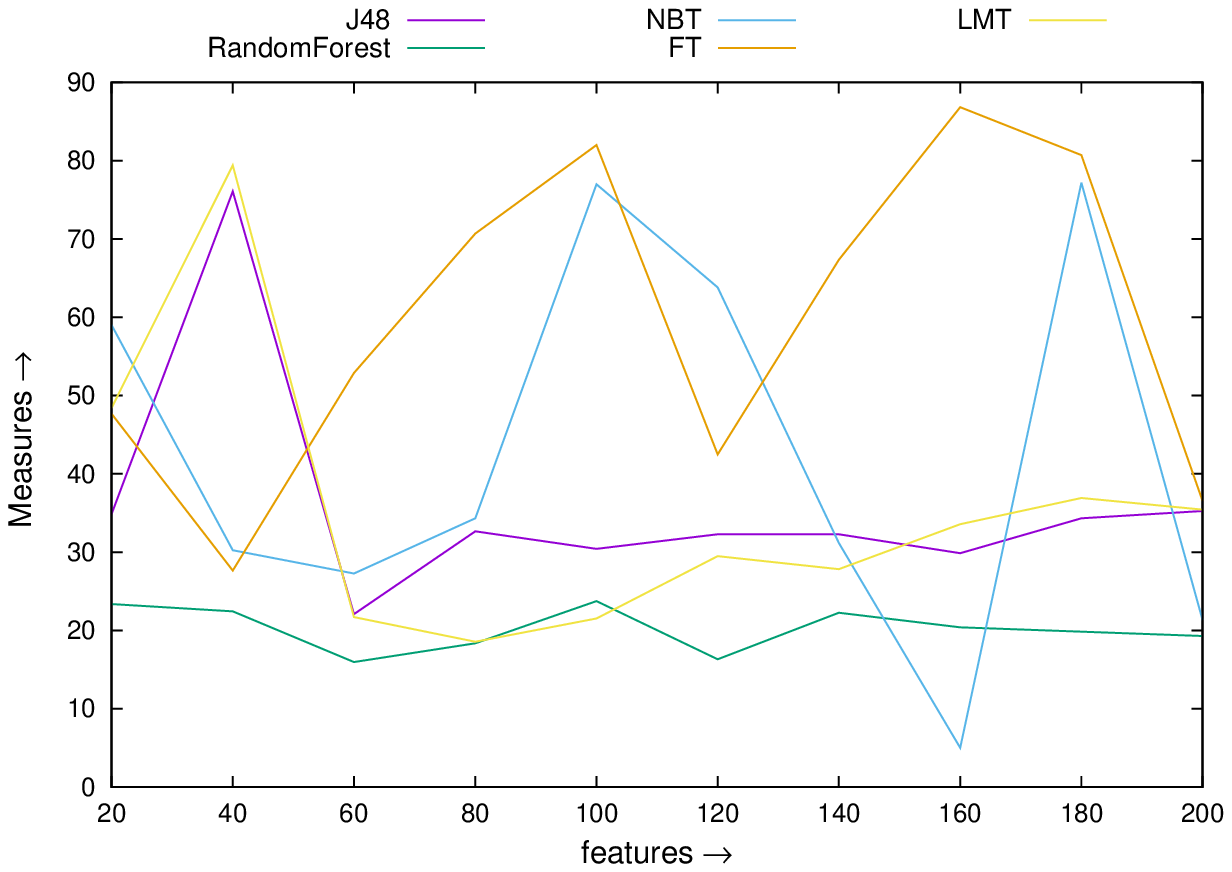} 
\caption{\small \sl True negatives obtained by selected five classifiers with different number of prominent features.} 
\label{tn}
    \end{minipage}   
\end{figure}

  \begin{figure}[!htb] 
\centering 
\begin{minipage}[t]{0.49\textwidth}
 \centering
        \includegraphics[width=1\linewidth, height=0.3\textheight]{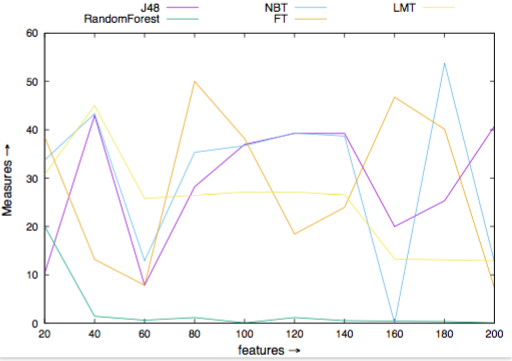} 
\caption{\small \sl False negatives obtained by selected five classifiers with different number of prominent features.} 
\label{fn}
  \end{minipage}  
\begin{minipage}[t]{0.49\textwidth}
       \centering
        \includegraphics[width=1\linewidth, height=0.3\textheight]{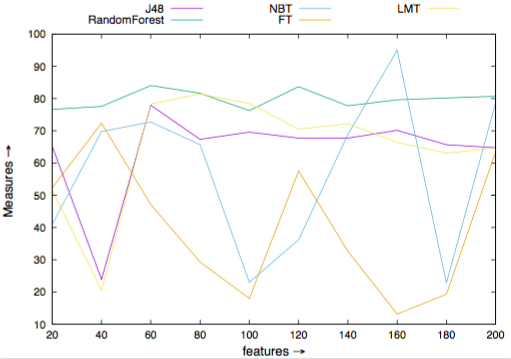} 
\caption{\small \sl False positives obtained by selected five classifiers with different number of prominent features.} 
\label{fp}
    \end{minipage}   
\end{figure}
  
\section{Result analysis}\label{sec:Result analysis}
The five selected classifiers are analysed by applying supervised machine learning technique with K-fold cross validation for k = 10. For the analysis, we first obtained the top 200 promising features (algorithm \ref{algo:FS}). The accuracy of the classifiers is obtained by varying the promising features and is measured by the equation
\begin{equation}
\text{Accuracy} = \frac{TP+TN} {TP + FN + TN + FP} \times 100
\end{equation}
\noindent where, \\                                                                          
	\noindent 	
		$TP \longrightarrow $ True positive, the number of malware apps correctly classified. \\
		$FN \longrightarrow $ False negative, the number of malware apps incorrectly classified. \\
		$TN \longrightarrow $ True negative, the number of benign apps correctly classified.\\
		$FP \longrightarrow $ False positives, the number of benign apps incorrectly classified.\\
			
The performance of the classifier has been studied by taking 20\% of available data (not used for training) with 20-200 best features, incrementing 20 features at each step and the result obtained are shown in figure.  \ref{fig:accuracy}.
From the analysis, the best accuracy is obtained by FT, Random forest, J48, LMT and NBT is approximately  $79.27$, $74.95$, $71.73$, $70.51$ and $68.87$ percent (figure \ref{ba}). Among these classifiers the least fluctuation in the accuracy by varying the features is observed in Random forest.
Figure \ref{tp} shows the TPR (malware detection rate) of all five classifiers with a different number of features. We found that the RF gives maximum TPR with least fluctuation compared to other classifiers. 

Figure ~\ref{tn} shows the TNR (benign detection rate) for all five classifiers with a different number of features. Here with some exception, we observed that FT detected the benign better than the other classifiers with a different number of features.
Figure \ref{fn} shows the false negatives of all selected classifier, in which compared to other classifiers the RF is good and also fluctuation is least with the number of features.
Figure \ref{fp} shows the false positives of the analysed classifiers and here we observed that all the five classifier does not give a good result, hence very much affects the final accuracy. However, although the false negative of RF is not as par but the fluctuation with the number of features is least compared to other classifiers. 

\section{Conclusion}\label{sec:Conclusion}

The threat/attack from the malicious apps in android devices are now never seen at before levels, as millions of android apps are available officially (google play store) and unofficially. Some of these available apps may be malicious, hence these devices are very much vulnerable to cyber threat/attack. The consequence will be devastating if in time counter-measures are not developed. 
Therefore, in this paper, we investigated five classifier FT, Random forest, J48, LMT and NBT for the detection of malicious apps. We found that among the studied classifiers, FT is the best classifiers and detect the malware with $\sim79.27\%$ accuracy. However, true positives i.e. malware detection rate is highest ($\sim99.91\%$) by RF and fluctuate least with the different number of prominent features  compared to other studied classifiers, which is better than BooJoong et. al., F-measure (98\%) \cite{kang2016n}. The analysis shows that overall accuracy is majorly affected by the false positives of the classifier. Hence in future more detail study are required to decrease the false positive and negative ratio for overall good accuracy and in this direction work is in progress, showing impressive results.

\section*{Acknowledgments}\label{sec:Acknowledgments}

Mr. Ashu Sharma is thankful to BITS, Pilani, K.K. Birla Goa Campus for the support to carry out his work through Ph.D. scholarship No. Ph603226/Jul. 2012/01.

\bibliography{Android}{}
\bibliographystyle{splncs03}

\end{document}